\documentclass[amsmath,amssymb,aps,prb,twocolumn,floatfix,unsortedaddress,superscriptaddress,reprint, longbibliography]{revtex4-2}
\usepackage{graphicx}
\usepackage[mathscr]{eucal}
\usepackage{bm}
\usepackage{textcomp,color}
\usepackage[colorlinks=true,urlcolor=blue,citecolor=blue,linkcolor=blue,bookmarks=false, pdfstartview={FitH}]{hyperref}
\usepackage[capitalize]{cleveref} 
\usepackage{subcaption}
\usepackage{float}
\usepackage{ragged2e}
\usepackage[normalem]{ulem} 

\newcommand{\nn}{\nonumber}

\newcommand{\msr}{\mathscr}

\newcommand{\beq}{\begin{equation}}
\newcommand{\eeq}{\end{equation}}
\newcommand{\bea}{\begin{eqnarray}}
\newcommand{\eea}{\end{eqnarray}}
\newcommand{\bse}{\begin{subequations}}
\newcommand{\ese}{\end{subequations}}
\newcommand{\bwt}{\begin{widetext}}
\newcommand{\ewt}{\end{widetext}}

\newcommand{\ua}{\uparrow}
\newcommand{\da}{\downarrow}

\begin{document}

\title{Electronic Raman response of a superconductor across a time reversal symmetry breaking phase transition}

\author{Surajit Sarkar}
\affiliation{Department of Physics, Concordia University, Montreal, QC H4B 1R6, Canada}
\author{Saurabh Maiti}
\affiliation{Department of Physics, Concordia University, Montreal, QC H4B 1R6, Canada}
\affiliation{Centre for Research in Molecular Modelling, Concordia University, Montreal, QC H4B 1R6, Canada}
\date{\today}

\begin{abstract}
Polarization-resolved electronic Raman spectroscopy is an important experimental tool to investigate collective excitations in superconductors. 
In this work, we present a general theory that allows us to study the evolution of all Raman active collective modes in multiple symmetry channels across a time-reversal symmetry (TRS) breaking superconducting transition. This comprehensive approach reveals that multiple modes belonging to different symmetry channels show a tendency to soften, even when the interactions in the subleading channel are held constant. This indicates an increased competition induced by the proximity to the TRS breaking transition. The entry into the TRS broken phase is marked by the introduction of an additional mode into the gap in multiple symmetry channels. These new modes have a phase character complementary to the ones that are already present. Even though all the modes in the TRS broken phase acquire an amplitude character, we explicitly demonstrate that the coupling to the Raman probe is exclusively through the phase sector. We demonstrate that the Raman spectrum collected in lower symmetry channels shows a selective sensitivity to the sign of the ground state order parameters and the sign of the interband interactions. Finally, we demonstrate the applicability of an interaction induced selection rule that clearly explains the spectral weights of various modes in various irreps, including the possible of ``dark" Leggett and Bardasis-Schrieffer modes.
%
%
\end{abstract}

\maketitle

\section{Introduction}\label{Sec:Introduction}
The use of electronic Raman scattering (eRS) to study the superconducting phase has a long history dating back to the study of 2H-NbSe$_2$ \cite{Sooryakumar1980} where superconductivity coexisted with a charge-density-wave. The interest in those days focused on understanding the coupling of phonon modes (either from the lattice of ions or from electrons as in charge-density-wave) to superconductivity\cite{Balseiro1980,Littlewood1982,Klein1982} as it was assumed that the cooper pair excitations, which were argued to be small\cite{Abrikosov1961}, would have a weak strength in the eRS spectra. After it was demonstrated (in A15 compounds) that these excitations could have observable spectra weight without the aid of phonons\cite{Dierker1983}, eRS was used to find the $T_c$, the size, temperature evolution, and even symmetry of the order parameter\cite{Hackl1983,Devereaux1994,Devereaux_prb95}. Many other uses of eRS to explore quantum phenomena in materials that are associated with superconductivity followed\cite{Devereaux_Rev_MP2007}. While phonons and magnons were the early (extrinsic) collective modes of interest to eRS, a purely electronic collective mode that was observed for the first time in MgB$_2$\cite{Blumberg_prl07} paved the way for eRS to be used to explore the nature and type of electronic correlations. Indeed, in the context of superconductivity, eRS was subsequently used to affirm the role of spin fluctuations in the pairing mechanism in the Fe-based superconductors\cite{maiti2016,Bohm2018} and identifying higher symmetry charge fluctuations\cite{Thorsmolle2016}. 

These developments, along with the fact that the same eRS experiment can provide access to excitations of the system in different irreducible representations (irreps) of the lattice simply by selecting the polarization of incoming and scattered light, encouraged both experimentalists and theorists to explore the role of correlations in the various quantum phases of matter. For example, all features of the eRS spectrum in the ${\rm A_{1g}}$ irrep for MgB$_2$, which is a 2 band system, was successfully explained by theory\cite{Klein2010,Cea_prb16}. However, subsequent theoretical studies for other materials suffered from limitations that prevented one from convincingly interpreting all features of eRS data. These limitations include the fact that most data from multiband systems were fitted using models for 1 band systems or systems idealized to have identical gaps and that the models were limited to one specific irrep at a time\cite{Scalapino2009,Chubukov2009,Khodas2014}. These were limitations because they never explained the simultaneous presence of significant spectral weight in both the collective mode and the $2\Delta$-threshold of Cooper-pair excitations, which is a common feature in many experiments. One had to invoke presence of independent bands to explain such features which is unrealistic in coupled multiband systems. This issue was soon addressed in a work\cite{maiti_prb17} that presented a general microscopic formalism to model a multiband eRS response with effects of electronic correlations in any symmetry channel for any pairing symmetry. This theoretical development allowed one to study the properties of any microscopic model with any number of bands based on a Fermi liquid normal state. One such work recently demonstrated the existence of additional interaction induced selection rules for the coupling of electronic collective modes\cite{BenekLins2023} to eRS that is applicable in all irreps. 

In this work we demonstrate another use of the general formalism and the interaction induced selection rule applied to the eRS response from a Time-Reversal symmetry (TRS) breaking superconductor and discuss the spectrum in various irreps across the phase boundary of a TRS to TRS breaking phase transition. In particular, we address questions about the number of collective modes across this transition, their energies, their character (what is the physical quantity that is fluctuating), their spectral weights in eRS, and if there are any characteristic changes between the two phases or across the transition. Our model even allows us to investigate the eRS spectra in different irreps. The motivation behind this type of study stems from the fact that TRS breaking superconductivity has attracted a lot of attention in recent years due to its application in topological quantum computation and other quantum devices\cite{Nayak_rmp08, Qi_rmp11}. Typical means of detecting TRS breaking are through the muon spin relaxation and the polar Kerr effect measurements. Indeed these tools were used to detect TRS breaking in UPt$_3$~\cite{Luke_prl93, Schemm_science14}, Sr$_2$RuO$_4$~\cite{Luke_nat98,xia_prl06}, URu$_2$Si$_2$~\cite{Schemm_prb15}, and in K-doped BaFe$_2$As$_2$\cite{Grinenkoprb17,Grinenko_nat20,Grinenko_nat21}. There are more predictions for TRS breaking state that are awaiting experimental affirmation such as for doped Graphene\cite{Nandkishore2012} and Moiré hetero-structures~\cite{balents_moire_prl18, yang_moire_prl18}. While detecting TRS breaking was possible, the order parameter is still debated in many of these materials\cite{Schemm_science14,Bhattacharya2023}. It is therefore beneficial to look for other probes that could provide some details about the order parameter structure, while still possibly detecting the TRS breaking transition. This work provides the initial steps toward modelling the eRS from such systems by studying $s+is$ state that have been identified in K-doped BaFe$_2$As$_2$.

The collective modes of a TRS breaking superconductor are partially known and are expected to couple to the eRS spectrum in general. However, the response itself has never been modelled microscopically, presumably because of a lack of theoretical progress. With our model, we can study the behavior of all the collective modes in the amplitude, phase, and density sectors, all the way across a phase boundary, and also model their spectral weights in various irreps of the eRS. We show that this can allow us to infer the phase of the order parameter, the attractive or repulsive nature of interactions, and even the phase boundary of the transition. There are many previously unacknowledged results about the collective mode spectrum and their coupling to eRS in such a system which we will spell out in the next section.

The rest of the text is organized as follows. In Sec. \ref{Sec:History} we present some historical discussion of the collective modes of a superconductor and their expected spectral weights and then summarize our main results in the context of this historical background. In Sec. \ref{Sec:low-energy-model} we summarize the multiband model for eRS in singlet superconductors, and provide some general considerations of the nature of collective modes and how they couple to the Raman probe. In Sec. \ref{Sec:result} we present a toy 3-band model that has a TRS to TRS breaking phase boundary. In Sec. \ref{Sec:A1g} we apply our general multiband theory across the entire phase diagram and discuss the eRS spectrum in the ${\rm A_{1g}}$ irrep. In Sec. \ref{Sec:B1g} we discuss the eRS spectrum in the ${\rm B_{1g}}$ irrep.  In Sec. \ref{Sec:Conclusion} we present our conclusions and some future directions.

\section{A brief history and main results}\label{Sec:History}
\paragraph*{Collective modes and their spectral weights:} The collective modes in a superconductor correspond to coherent fluctuations of the order parameter (also referred to as the excitations of the Cooper pair). In a single band superconductor (we only focus on clean systems at $T=0$) there are two collective modes. One corresponds to the coherent fluctuations of the order parameter strength (amplitude): the massive Anderson-Higgs(AH) mode\cite{anderson_gauge58,Vaks62}. This mode does not directly couple to spectroscopic probes in the linear regime (unless mediated by phonons\cite{Balseiro1980,Littlewood1982,Klein1982,Tutto1992} or external supercurrent\cite{shimano_prl19}). The other mode corresponds to the coherent fluctuations of the phase of the order parameter: massless the Bogoliubov-Anderson-Goldstone (BAG) mode\cite{anderson_gauge58,anderson1958,Bogoljubov58}. Modes in the phase sector couple to density fluctuations which in turn directly couple to photons. The BAG mode, although expected due to the spontaneous breaking of U(1) symmetry, actually is renormalized by the Coulomb interaction to the plasma frequency and hence not accessible at low energies\cite{anderson63}. Even if they were probed at higher energies, the spectral weight associated with this mode would be $\propto q^2\rightarrow 0$\cite{Vaks62,Klein_prb84}, where $q$ is the momentum transferred by photons, rendering it undetectable. If there were competing Cooper channel interactions in irreps other than the ground state one, then one could expect additional massive collective modes both in phase and amplitude sectors. These phase sector collective modes, which couple to density fluctuations and hence to photons, are called the Bardasis-Schrieffer (BaSh) modes\cite{Bardasis61,Vaks62,Scalapino2009}. While the AH and BAG modes are present but never visible in superconductors, the BaSh mode, when present, are expected to be visible. The presence of BaSh mode signals a strong competition from a subleading irrep in the Cooper channel.

In a two-band superconductor, the collective excitations include additional massive collective modes. In the phase sector, these are called the Leggett modes\cite{Leggett66}. They correspond to the out-of-phase phase fluctuations of the order parameters in two bands, as opposed to the in-phase BAG mode which remains massless. Similarly, more modes are expected in higher irreps if there is competition. None of the modes except the BAG are normalized by the Coulomb interaction due to either a symmetry decoupling (from belonging to different angular momentum channels) or due to the fact that the long-range Coulomb interaction is blind to the intra-unit cell part of the fluctuations that exist in the multiband case\cite{Abrikosov1974}. All these additional modes in the phase sector are expected to have finite spectral weight, while the amplitude sector remains inaccessible.

In TRS broken superconductors, however, the amplitude and the phase sector fluctuations are coupled. For a TRS breaking $s+is$ state, the expected collective mode spectrum across this phase transition is such that a Leggett mode softens and bounces back\cite{Lin_prl12,maitiprb13,Marciani_prb13}. For an $s+id$ transition, there is a mixed symmetry BaSh mode that is expected to soften and bounce back\cite{maiti_prb15}. There are also other predictions for the existence of collective modes in the TRS broken phases \cite{Balatsky_prl2000,Lee_prl09,Poniatowski_comm22}. However, their existence does not guarantee their coupling to a probe. Addressing this was one of the main motivations for this work. Specifically, there are open questions as to if the modes in the phase sector are always expected to couple to eRS; if the modes in the TRS broken state that include amplitude fluctuations enable an equilibrium coupling of photons to the amplitude sector; and what are the spectral weights of various modes in both the phases and across the transition.

\paragraph*{Summary of main findings:} We worked with the $s$ to $s+is$ transition that requires at least three bands microscopically. This allows us to use constant gaps which keeps the calculations analytically tractable. The general results for the collective modes are as follows. (i) In the $\rm A_{1g}$ irrep there is the expected Leggett mode that softens when approaching the phase boundary from the TRS phase, and it bounces back in the TRS broken phase. Surprisingly there is a similar tendency in the $\rm B_{1g}$ irrep with the BaSh mode, which happens without altering the $\rm B_{1g}$ interactions. This suggests that the approach to the TRS breaking boundary is marked by competition between $s+is$ and $s+id$ states. (ii) In the TRS broken phase, there is an additional collective mode that is induced near the pair-breaking threshold in both irreps. Both the bounced-back mode and the additional mode have mixed amplitude and phase character of the fluctuations. However, near the TRS breaking transition one of the modes has predominantly an amplitude character while the other has a phase character. The phase part of the fluctuations of these two modes have complementary characters: if one is in-phase, the other is out-of-phase. (iii) In the $\rm B_{1g}$ sector we show that depending on the sign of the interband interactions either the in-phase or out-of-phase mode could be the low energy mode. This is helpful as it indicates the phase characteristic of the competing $d$-wave state in terms of it being of the $d^{++}$ or the $d^{+-}$ type.

The general results for the manifestation of collective modes in eRS are far more interesting which can be explicitly inferred from the formulas we derive but, in many cases, can also be deduced from the interaction induced selectivity of modes outlined in Ref. \cite{BenekLins2023} (cf. \ref{Sec:A1g} and \ref{Sec:B1g} below). These findings are as follows. (i) Despite the mixing of the phase and amplitude fluctuations, the coupling to eRS comes exclusively from the phase fluctuations of the collective mode. This is true in all irreps. (ii) In both the TRS and TRS broken phases, the ${\rm A_{1g}}$ response is not sensitive to the sign of the gaps in the ground state, but the $\rm B_{1g}$ response is and this could serve as means to detect sign changes of the order parameter. (iii) There is a possibility of ``dark" Leggett mode and BaSh mode that prevents these modes from coupling to eRS. Thus, a phase transition could be triggered without a visible softening of a mode. (iv) The spectral weight near the pair-breaking continuum does not display any characteristic change across the transition. However, they are characteristically different deep inside and deep outside the TRS broken phase.

The knowledge of the evolution of the character and spectral weight of the modes in different irreps provides a clear picture of the low-energy sector near a TRS breaking transition and could be used to even detect such transitions. In the remainder of the text, we will substantiate the above statements with detailed explanations and discussions.

\section{The Raman response in a multiband superconductor}\label{Sec:low-energy-model}
Consider a multiband system $\mathscr{H} = \msr{H}_{\rm{MF}} +\msr{H}_{\rm{residual}}$. Here $\msr{H}_{\rm{MF}}$ is the mean-field part of the Hamiltonian in the singlet channel given by $\msr{H}_{\rm{MF}}=\sum_{\alpha,\vec k}\Psi^\dag_\alpha(\vec k)\hat{\msr{E}}_\alpha(\vec k)\Psi_\alpha(\vec k)$, where $\hat{\msr{E}}_\alpha(\vec k)=\varepsilon_{\alpha,\vec k}\hat\sigma_3- \Delta^R_{\alpha,\vec k}\hat\sigma_1 +\Delta^I_{\alpha,\vec k}\hat\sigma_2$, and $\Psi_\alpha(\vec k)=\left(\hat c_{\alpha,\vec k,\ua}~\hat c^\dag_{\alpha,-\vec k,\da}\right)$. The index $\alpha\in\{a, b,...\}$ represents the bands, $\hat{\sigma}_0$ is a $2\times2$ identity matrix, $\hat\sigma_i, i\in\{1,2,3\}$ are the Pauli matrices in the particle-hole space. Further, $\varepsilon_{\alpha,\vec k}$ is the dispersion of $\alpha^{\rm th}$ band and $\hat c_{\alpha\vec k s}$ denotes the annihilation operator for the quantum state in band $\alpha$, momentum $\vec k$ and spin $s$. Finally, $\Delta_{\alpha,\vec k}\equiv \Delta^R_{\alpha,\vec k}+i\Delta^I_{\alpha,\vec k}$ is the mean-field order parameter. $\msr{H}_{\rm{residual}}$ denotes the residual interactions in the Cooper channel that correspond to momentum transfers $\vec q\neq 0$, where $\vec q=0$ is the momentum transfer channel in which the condensation took place ~\cite{maiti_prb10, maiti_prb17}. These interactions include density-density interaction within the same band as well as density-density, exchange, and pair-hopping interactions between fermions from different bands. Of these interactions, however, what is relevant for the superconductivity problem is just the Cooper channel projection of the interactions (see Ref. \cite{maiti_prb17}) which we model as $V^{\rm pp}_{\alpha\beta}$ between two bands $\alpha$ and $\beta$. In what follows, to ensure analytical tractability, we shall work at temperature $T=0$ with $\vec k$-independent interactions, which results in $\vec k$-independent order parameters $\Delta_{\alpha}=\Delta^R_\alpha+i\Delta^I_\alpha$ which satisfy the self-consistency equations
\bea\label{eq:self_con1}
\Delta_{\alpha}=-\sum_{\beta}V_{\alpha\beta}^{\rm pp}\Delta_\beta \nu_F^\beta l_\beta,
\eea
where $\nu_F^\beta$ is the density of states at the Fermi level of band $\beta$, and $l_\beta\equiv\ln(2\Lambda/|\Delta_\beta|)$, where $\Lambda$ is some cut-off associated with the pairing mechanism.

It was shown in Ref.~\cite{maiti_prb17} that for $\vec q\rightarrow 0$ (the momentum transferred by light to the superconductor) the long-range Coulomb interaction did not affect the response. This allows us to unify the treatment of the Raman response in all irreps $r$ of the lattice which is computed as Im[$\chi_r(\Omega)$], where $\chi_r(\Omega)$ is obtained from the analytic continuation of \cite{maiti_prb17}:
\bea\label{eq:response_general}
\chi_r(Q)&=&-[c_r]^T[[\Pi]][\Gamma_r],~~\text{where}\nn\\
~[c_r]&=&(0,0,\gamma_r^a, 0,0,\gamma_r^b,...)^T,\nn\\
~[[\Pi]]&=&{\rm Diag}([\Pi^a],[\Pi^b],...),\nn\\
~[\Pi^\alpha]&=&\Pi^\alpha_{ij}(Q),~\text{for}~i,j\in\{1,2,3\},\nn\\
\Pi^{\alpha}_{ij}(Q)&\equiv&\int_K \text{Tr}[\hat\sigma_i\hat{\msr{G}}_\alpha(K)\hat\sigma_j\hat{\msr{G}}_\alpha(K+Q)],\nn\\
~[\Gamma_r]&=&[\mathcal{R}]^{-1}[c_r],\nn\\
~[\mathcal{R}]&=&\mathbb{I}_{3n\times3n}+\frac12[[V^{\rm pp}_r]][[\Pi^{\rm pp}]],\nn\\
~[[V^{\rm pp}_r]]&=&V^{\rm pp}_{r,\alpha\beta}\otimes\mathbb{I}_{3\times3},\nn\\
~[[\Pi^{\rm pp}]]&=&{\rm Diag}([\Pi^{\rm pp,a}],[\Pi^{\rm pp,b}],...),\nn\\
~[\Pi^{\rm pp,\alpha}]&=&-\begin{pmatrix}
\Pi^\alpha_{11}&\Pi^\alpha_{12}&\Pi^\alpha_{13}\\
\Pi^\alpha_{21}&\Pi^\alpha_{22}&\Pi^\alpha_{23}\\
0&0&0
\end{pmatrix}.
\eea
where $\gamma^\alpha_r$ is the projection of the effective mass vertex of band $\alpha$ onto the $r^{\rm th}$ irrep, $n$ is the number of bands,  $K\equiv(i\omega_n,\vec k)$, $Q\equiv(i\Omega_m,\vec q)$,   $\int_K\equiv T\sum_n\int d^dk/(2\pi)^d$ (where $d$ is the spatial dimension), and $\msr{G}_\alpha(i\omega_n,\vec k)$ is the Green’s function:
\bea
\hat{\msr{G}}_\alpha(i\omega_n,\vec k)=\Big[i\omega_n\hat{\sigma}_0-\hat{\msr{E}}_\alpha(\vec k)\Big]^{-1}.
\eea
$V^{\rm pp}_{r,\alpha\beta}$ is the projection of the Cooper channel interaction onto the $r^{\rm th}$ irrep. In the above expression for the Raman response, we have ignored the contributions from interactions in the particle-hole channel. See Ref \cite{maiti_prb17} for its inclusion. The various correlation functions $\Pi^\alpha_{ij}$ are evaluated as
\bea\label{eq:pis}
\frac{\Pi^\alpha_{11}(\Omega)}{2\nu_F^\alpha}&=&-l^\alpha-\left[\left(\frac{\Omega}{2|\Delta_\alpha|}\right)^2-\left(\frac{\Delta_\alpha^R}{|\Delta_\alpha|}\right)^2\right]\mathscr{I}_\alpha(\Omega),\nn\\
\frac{\Pi^\alpha_{22}(\Omega)}{2\nu_F^\alpha}&=&-l^\alpha-\left[\left(\frac{\Omega}{2|\Delta_\alpha|}\right)^2-\left(\frac{\Delta_\alpha^I}{|\Delta_\alpha|}\right)^2\right]\mathscr{I}_\alpha(\Omega),\nn\\
\frac{\Pi^\alpha_{33}(\Omega)}{2\nu_F^\alpha}&=&-\mathscr{I}_\alpha(\Omega),\nn\\
\frac{\Pi^\alpha_{12}(\Omega)}{2\nu_F^\alpha}&=&-\left(\frac{\Delta_\alpha^R\Delta_\alpha^I}{|\Delta_\alpha|^2}\right)\mathscr{I}_\alpha(\Omega)=\Pi^\alpha_{21}(\Omega),\nn\\
\frac{\Pi^\alpha_{13}(\Omega)}{2\nu_F^\alpha}&=&-i\left(\frac{\Omega\Delta_\alpha^I}{2|\Delta_\alpha|^2}\right)\mathscr{I}_\alpha(\Omega)=-\Pi^\alpha_{31}(\Omega),\nn\\
\frac{\Pi^\alpha_{23}(\Omega)}{2\nu_F^\alpha}&=&-i\left(\frac{\Omega\Delta_\alpha^R}{2|\Delta_\alpha|^2}\right)\mathscr{I}_\alpha(\Omega)=-\Pi^\alpha_{32}(\Omega),
\eea
where 
\bea
\mathscr{I}_\alpha(\Omega)=\frac{\sin^{-1}(\Omega/2|\Delta_\alpha|)}{(\Omega/2|\Delta_\alpha|)\sqrt{1-(\Omega/2|\Delta_\alpha|)^2}}.
\eea
In $\Pi^\alpha_{ij}$, $i=1$ is associated with the amplitude sector fluctuations, $i=2$ is associated with the phase sector fluctuations, and $i=3$ is the density sector fluctuations. Technically, $[\Pi^{\rm pp}]$ is not given in terms of $\Pi^\alpha_{ij}$ but a quantity whose integrand is the same as $\Pi^\alpha_{ij}$ but dressed with normalized angular form factors of the $r^{\rm th}$ irrep. Since our $\Delta_\alpha$ are all $\vec k$-independent, the angular integration always yields unity leading to the above form of the equations. The inclusion of $\Delta^I_\alpha$ is the novel aspect of this formalism compared to the previous works\cite{maiti_prb17}.

\subsection{General considerations}\label{Sec:GenCon}
The resonances in the Raman response are captured as poles of $[\Gamma_r]$ which stem from the zeroes of Det[$\mathcal{R}$], where $\mathcal{R}$ is a $3n\times3n$ matrix for the case of $n$ bands. The 3 arises from the amplitude, phase, and density degrees of freedom in the superconductor. We can estimate the number of zeroes in the following manner. Focus first on the region $\Omega\ll2|\Delta|$ where the frequency dependence is such that $\Pi^\alpha_{11}$ and $\Pi^\alpha_{22}$ $\sim\Omega^2$ each, $\Pi^\alpha_{33},\Pi^\alpha_{12}$ are independent of $\Omega$, and $\Pi^\alpha_{13},\Pi^\alpha_{23}$ are linear in $\Omega$ [see Eq. (\ref{eq:pis})] but appear as a product of themselves. Thus, ${\rm Det}[\mathcal{R}]=0$ is at best $2^{\rm nd}$ order polynomial in $\Omega^2$ per band or $2n^{\rm th}$ degree polynomial overall. It is not of degree $3n$ as $\Pi^\alpha_{33}$ has no $\Omega^2$ term. Next, due to the global U(1) symmetry, the order parameter at the mean-field level is usually chosen to be real if TRS is preserved, implying $\Delta^I_\alpha=0$ which leads to $\Pi^\alpha_{12}=0=\Pi^\alpha_{13}$ decoupling the amplitude sector from the phase and density sector. This decouples the determinant into products of two $n^{\rm th}$ order polynomials in $\Omega^2$. This factorization separates the amplitude ($i=1$) and the phase ($i=2$) sector modes such that there are $n$ of each. Only the phase sector modes are Raman-active (i.e. have finite spectral weight) as this is the sector that couples, via non-zero $\Pi^{\alpha}_{23}$, to the effective mass vertex that involves the density vertex with $i=3$, which in turn couples to photons. Observe also that if we had chosen the gap to be purely imaginary, then $\Delta^R_\alpha=0$ leading to $\Pi^\alpha_{23}=0=\Pi^\alpha_{21}$. Then $\Delta^I$ becomes the amplitude sector and is still decoupled from the phase ($i=1$) and density ($i=3$) sectors. 

If we did not assume anything about the ground state phase, all the components above would be formally coupled. However, note that the coupling of the two sectors via $\Pi^\alpha_{12},\Pi^\alpha_{13}$ terms don't add any additional powers of $\Omega^2$ leading to the same total mode count of $2n$. Further, the self-consistency equations [which knows about the U(1) symmetry] will cause any observable result to be only dependent on $|\Delta_\alpha|$ as can be explicitly checked by using Eq. (\ref{eq:pis}) in Eq. (\ref{eq:response_general}). This means that the modes would still decouple into $n$ phase sector modes and $n$ amplitude sector modes, with only the phase sector modes being potentially Raman-active. This check ensures the U(1) gauge invariance of our formalism. If the ground state breaks TRS then the result would depend on the complex phase of the order parameter, but this does not affect the mode count as we already saw above that the inclusion of all the terms of $\Pi^\alpha_{ij}$ don't add any additional powers of $\Omega^2$. What is different in this case is that the modes would have mixed amplitude and phase character, as was noted in previous works\cite{Lin_prl12,maitiprb13, Marciani_prb13}. In this case, however, the results will be sensitive to the phase of the order parameter $\phi$.

\subsection{The character of the modes}\label{Sec:Char}
In order to understand the nature of the modes that are Raman-active, let us note a very special property of  $\chi^\alpha_{22}\equiv\Pi^\alpha_{22}+2\nu_Fl_\alpha,\Pi^\alpha_{23},\Pi^\alpha_{32}$ and $\Pi^\alpha_{33}$: $\chi^\alpha_{22}\Pi^\alpha_{33}-\Pi_{32}\Pi_{23}=0$. This can be seen as ${\rm Det}[\Pi^{\alpha}_{ij}]=0$ with $\Pi^\alpha_{22}\rightarrow\chi^\alpha_{22}$, where $i,j\in\{2,3\}$. This property of the zero determinant is responsible for the fact that one of the phase sector collective modes in the channel $r={\rm A_{1g}}$ is always the massless BAG mode. This mode is removed from the low energy sector by the Coulomb interaction, but would not have shown up anyway due to its zero mass. Thus, only $n-1$ modes in the phase sector are Raman active for $r={\rm A_{1g}}$, which are the Leggett-type modes. In other irreps, we still have $n$ phase modes that are Raman-active and these are the BaSh-type modes. Note that only modes with $\Omega<2|\Delta_{\rm min}|$ (the minimum gap in the system)] will be long-lived. The modes in the continuum of the pair-breaking excitations will be damped. 

It is worth noting (see Appendix \ref{Sec:App1}) that even if we included $\Delta^I_\alpha\neq0$, we would have ${\rm Det}[\Pi^{\alpha}_{ij}]=0$ with $i,j\in\{1,2,3\}$ and $\Pi^\alpha_{11}\rightarrow\chi^\alpha_{11},\Pi^\alpha_{22}\rightarrow\chi^\alpha_{22}$ (the correlation functions without the $l_\alpha$ part), ensuring that we still have the Goldstone mode in the $r={\rm A_{1g}}$ channel. This aspect does not depend on whether TRS is preserved or broken and hence the Goldstone mode is always present and is massless (and is always removed by the Coulomb interaction). What changes in the TRS broken phase is that now all the $2n-1$ modes become Raman-active due to the mixing of the amplitude and phase contributions. What can also happen is that some modes from the continuum could be pushed into the gap, but this does not change the mode count.

To quantify the admixing of the amplitude and phase fluctuations for the various modes, we can follow the treatment in Refs. \cite{maitiprb13,maiti_prb15} where the dynamical form of the self-consistency equation is expanded to linear order in the fluctuating components: $\delta\Delta_\alpha\equiv\delta^R_\alpha+i\delta^I_\alpha$. The presence of non-trivial solutions to this equation for a given value of frequency would indicate that the system supports spontaneous fluctuations at that frequency. This maps the problem of finding collective modes to that of finding the eigenvalues and eigenvectors of the dynamical self-consistency equation of the fluctuating components. The eigenvalues give the collective mode frequencies while the eigenvectors give the character, i.e. the extent of admixture of phase and amplitude fluctuations, of the collective modes.

At a finite wavevector $q$ of the fluctuations, we would also need to account for variation in the density ($\delta\rho_\alpha$). The general eigenvalue problem would be given by:
\bea\label{eq:linResp}
2\delta^R_\alpha &=&\sum_{\beta}V^{\rm pp}_{\alpha\beta}\left[\Pi_{11}^\beta\delta^R_\beta-\Pi_{12}^\beta\delta^I_\beta+\Pi_{13}^\beta\delta\rho_\beta\right]\nn\\
-2\delta^I_\alpha &=&\sum_{\beta}V^{\rm pp}_{\alpha\beta}\left[\Pi_{21}^\beta\delta^R_\beta-\Pi_{22}^\beta\delta^I_\beta+\Pi_{23}^\beta\delta\rho_\beta\right]\nn\\
\delta\rho_\alpha&=&\sum_{\beta}V_{q}\left[\Pi_{31}^\beta\delta^R_\beta-\Pi_{32}^\beta\delta^I_\beta+\Pi_{33}^\beta\delta\rho_\beta\right],
\eea
where $V_q$ is the singular Coulomb interaction ($\sim 1/q$ in 2D and $\sim 1/q^2$ in 3D) and is the same for all bands. This can be recast into the form $[\mathcal{K}]~[\delta]=0$ where $[\delta]\equiv(\delta^R_a,-\delta^I_b,\delta\rho_a,\delta^R_b,-\delta^I_b,\delta\rho_b,...)^T$ and 
\bea\label{eq:matrixform}
[\mathcal{K}]&=&\mathbb{I}_{3n\times 3n}+\frac12[[V^{\rm pp}_{gs}]][[\Pi^{\rm pp}]]+\frac12[[V_{C}]][[\Pi^{\rm ph}]],\nn\\
\text{where,}\nn\\
~[[V_C]]&=&2V_q
\begin{pmatrix}
1&1&\cdots&1\\
1&1&\cdots&1\\
\vdots&\vdots&\ddots&\vdots\\
1&1&\cdots&1
\end{pmatrix}_{n\times n}\otimes\mathbb{I}_{3\times 3},\nn\\
~[[\Pi^{\rm ph]}]]&=&{\rm Diag([\Pi^{\rm ph,a}],[\Pi^{\rm ph,b}],...)},\nn\\
~[\Pi^{\rm ph,\alpha}]&=&\begin{pmatrix}
    0&0&0\\
    0&0&0\\
    \Pi^{\alpha}_{31}&\Pi^{\alpha}_{32}&\Pi^{\alpha}_{33}
\end{pmatrix}.
\eea
Here $V^{\rm pp}_{gs}$ denotes the pairing interaction in the ground state irrep as in Eq. (\ref{eq:response_general}). The $\mathcal{K}$ is the same matrix $\mathcal{R}$ that appears in the Raman response with $r=gs$. In the specific form in Eq. (\ref{eq:response_general}) we dropped the $[V^{\rm ph}]$ part (see Ref. \cite{maiti_prb17} for the full form) on grounds of studying the effect of only the Cooper channel interactions. In the above eigenvalue problem, the Coulomb interaction only becomes relevant at finite $q$. Since our focus is on the $q=0$ part, we will continue to ignore the coupling to the density sector to maintain consistency in our treatment (formally we are ignoring $1/V_C$ compared to $[[V^{\rm pp}_{gs}]^{-1}]_{\alpha\beta}$).

\begin{figure}[htb!]
\centering\captionsetup{justification=RaggedRight}
\includegraphics[width=\linewidth]{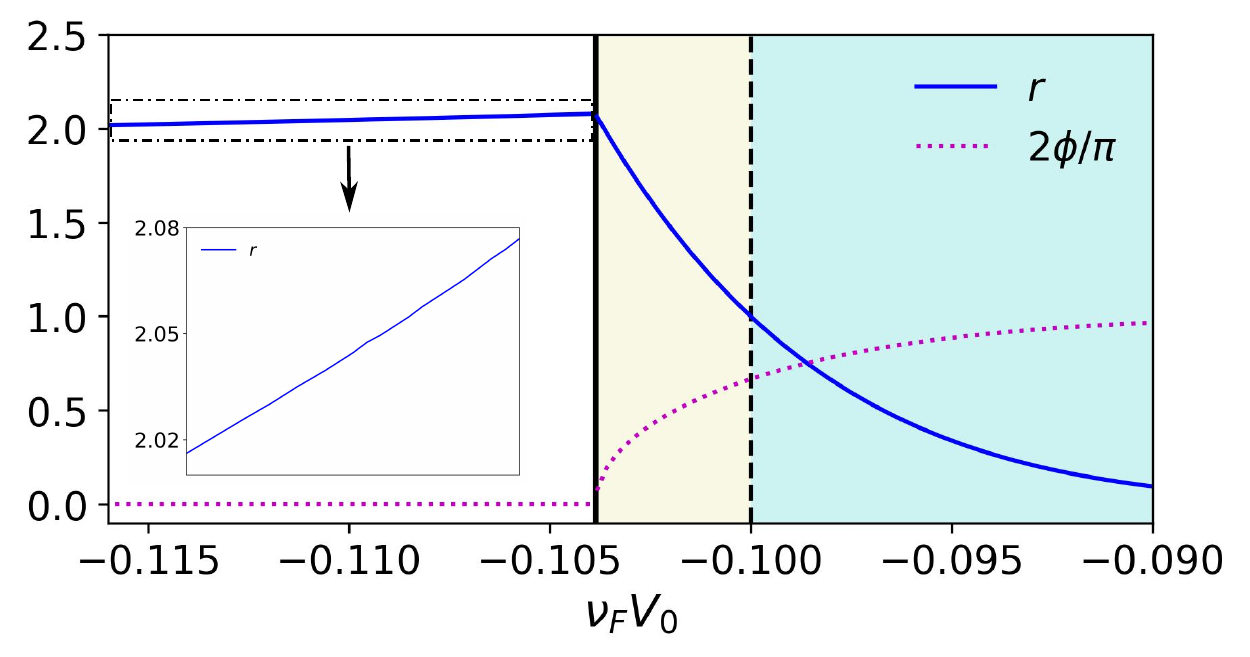}
\caption{The parameter $r=\Delta_0/\Delta_1$ w.r.t $\nu_FV_0$ across an $s^{++}\rightarrow s+is$ transition, which happens at $\nu_FV_0\approx-0.104$ (indicated by the solid vertical line). Here the phase $\phi$ (dotted curve) becomes non-trivial. The vertical dashed line shows the point where $r=1$ beyond which we label the state as being ``deep" in the TRS broken phase. The phase diagram for $s^{+-}\rightarrow s+is$ transition remains exactly the same but with $V_0\rightarrow-V_0$. Here $\nu_FV_1=0.1$ and $\Lambda=1.5\times10^4\Delta_1$. The inset is provided to highlight the variation in $r$ with $\nu_FV_0$.}
\label{phase-diag}
\end{figure}

Since the linear response matrix ($\mathcal{K}$) and the kernel of the Raman response ($\mathcal{R}$) are the same, the character of the collective modes, which are the zeroes of the determinant of $\mathcal{R}$ or $\mathcal{K}$, can be inferred by studying the corresponding eigenvector whose components represent the weights of the fluctuating terms of $\Delta_\alpha$. To categorize these fluctuations in terms of amplitude and phase note that $$\delta\Delta_\alpha\equiv\delta(|\Delta_\alpha|e^{i\phi_\alpha})\approx e^{i\phi_\alpha}\delta|\Delta_\alpha|+i\Delta_\alpha \delta\phi_\alpha+...$$
where the $...$ stand for higher order terms in $\delta|\Delta_\alpha|$ and $\delta\phi_\alpha$. Taking the real and imaginary parts of this we get the expressions for $\delta_\alpha^R$ and $\delta_\alpha^I$ which we also get from the eigenvectors of $[\mathcal K]$. Inverting this relation we arrive at the amplitude and phase parts to be:
\bea 
\delta|\Delta_{\alpha}|=\delta^R_{\alpha}\cos\phi_\alpha+\delta^I_{\alpha}\sin\phi_\alpha,\nn\\
|\Delta_\alpha|\delta\phi_{\alpha}=\delta^I_{\alpha}\cos\phi_\alpha-\delta^R_{\alpha}\sin\phi_\alpha.
\eea
We then define the amplitude and phase characters of a mode $m$ as:
\beq
C^A_m\equiv\sum_\alpha\left|\frac{\delta|\Delta_{\alpha,m}|}{|\Delta_\alpha|}\right|^2,~~
C^P_m\equiv\sum_\alpha\left|\delta\phi_{\alpha,m}\right|^2,
\eeq
which simply sums up the weights in the amplitude or phase sector from all the bands.

We will now apply this general formalism of finding collective modes and their character to the minimal model that demonstrates TRS breaking with $k$-independent interactions.

\begin{figure*}[htb!]
\centering\captionsetup{justification=RaggedRight}
\includegraphics[width=0.49\linewidth]{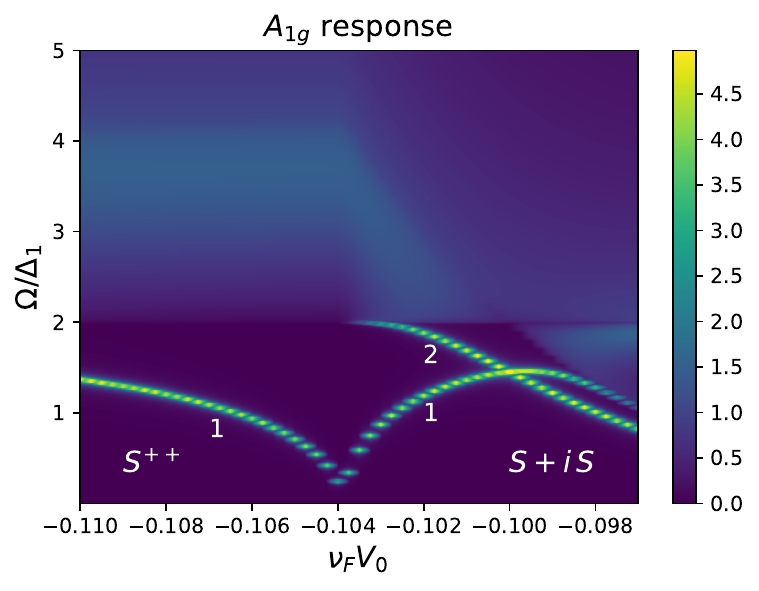}
\includegraphics[width=0.49\linewidth]{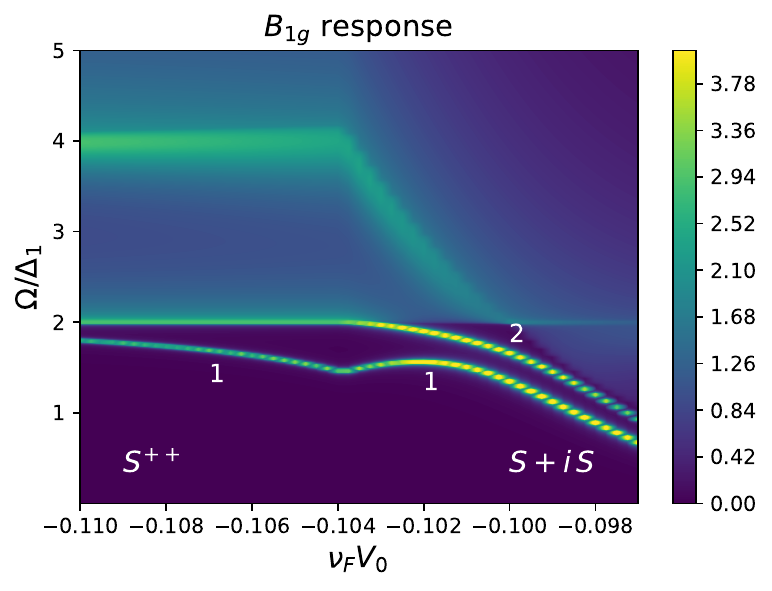}
\caption{The eRS response in the $\rm A_{1g}$ and $\rm B_{1g}$ irreps. The $\rm A_{1g}$ mode 1 softens at the $s^{++}\rightarrow s+is$ transition ($\nu_FV_0\approx-0.104$) while the $\rm B_{1g}$ mode 1 turn's back. In both cases, an additional mode 2 is induced from the $2\Delta$ threshold inside the TRS broken phase. The choice of Raman vertices in both cases is $\gamma^b_{r}=-\gamma^c_{r}$, $r\in\{\rm A_{1g}, B_{1g}\}$. For $\rm A_{1g}$, all parameters are the same as in Fig. \ref{phase-diag}. For $\rm B_{1g}$, the interband interaction elements are $\nu_F U_0=0.07, \nu_F U_1=0$. A fermion lifetime of $0.005\Delta_1$ was added to induce broadening. The color scale corresponds to $\log(1+\chi_{r}/\chi_0)$, where $\chi_0 \propto \nu_F(\gamma^a_r)^2$ is an arbitrary scale to remove the dimensions. This form was chosen to enhance the visibility of the $2\Delta$ region.}
\label{fig:Contour}
\end{figure*}

\begin{figure}[htb!]
\centering\captionsetup{justification=RaggedRight}
\includegraphics[width=\linewidth]{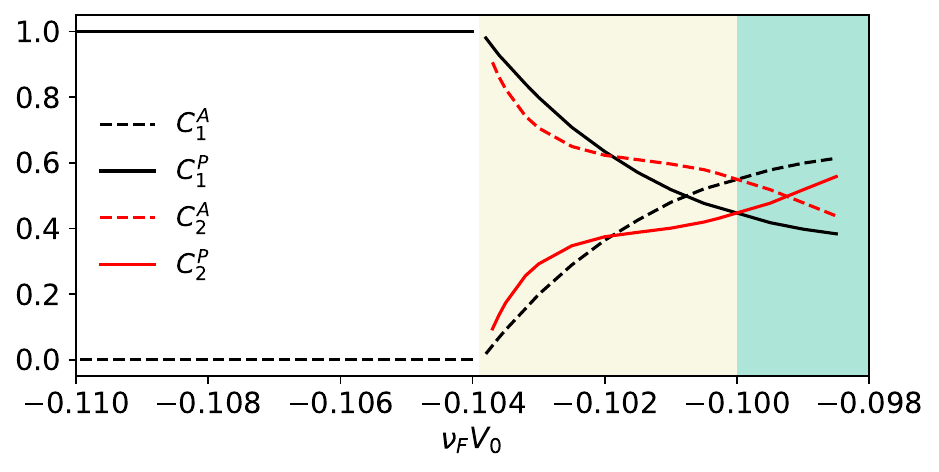}
\caption{The amplitude and phase character strengths $C^A_{m},C^P_m$, respectively, for mode $m$. In the TRS side, the mode is a pure phase one. In the TRS broken side, the characters mix and the phase-dominant mode 1 evolves into an amplitude-dominant one whereas the amplitude-dominant mode 2 evolves into a phase dominated one. All parameters are the same as in Fig. \ref{fig:Contour}.}
\label{character}
\end{figure}

\begin{figure*}[htb!]
\centering\captionsetup{justification=RaggedRight}
\includegraphics[width=1.0\linewidth]{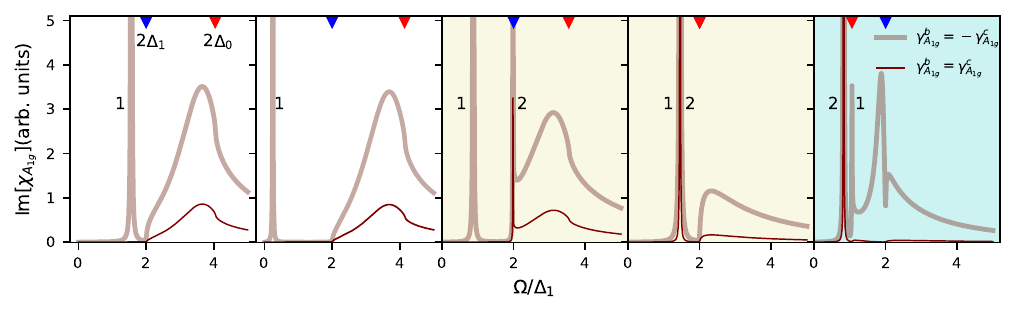}
\caption{The ${\rm A_{1g}}$ response across the $s^{++}\rightarrow s+is$ transition for different choices of Raman vertices. The five panels are for $\nu_FV_0\in\{ -0.114,-0.104, -0.103, -0.1, -0.097\}$ from left to right. The background color corresponds to that shown in Fig. \ref{phase-diag} with the white being the TRS phase, the yellow being the TRS broken phase and the blue being deep in the TRS broken phase. The labels 1 and 2 track the long-lived collective modes shown in Fig. \ref{fig:Contour} and the two inverted triangles at the top track the evolution of the gaps $\Delta_1$ (for bands $b,c$) and $\Delta_0$ (for band $a$). All parameters are the same as in Fig. \ref{fig:Contour}. Observe that mode 1 does not show up in eRS for $\gamma^b_{A_{1g}}=\gamma^c_{A_{1g}}$.}
\label{fig:A1g}
\end{figure*}

\section{Phase diagram of 3-band model with broken TRS}\label{Sec:result}
Consider a system with bands $a,b,c$. To keep the solution analytically tractable, we make bands $b$ and $c$ to be identical. This allows us to model the interband interactions in the Cooper channel as 
\beq\label{eq:interactions}
V^{\rm pp}_{\alpha\beta}\rightarrow\begin{pmatrix}
     0&V_0&V_0\\
     V_0&0&V_1\\
     V_0&V_1&0
\end{pmatrix}.
\eeq
Here we have ignored intra-band interactions as we are mostly interested in the new aspects due to the multiband nature of the pairing problem. Certainly, the formalism can be applied to any general interaction matrix at the cost of studying the effects of more parameters. The most general TRS broken state in the above model could then be parameterized as $\Delta_a=\Delta_0>0$, $\Delta_b=\Delta_1e^{i\phi}$, and $\Delta_c=\Delta_1e^{-i\phi}$ ($\Delta_1>0$) with the order parameters in band $b$ and $c$ differing only in phase\footnote{This model, where the bands $b,c$ were seen as pockets was used in Ref. \cite{maitiprb13}.}. Further, the overall scale of the order parameters will be controlled by the cut-off and hence a relevant parameter would be $r\equiv\Delta_0/\Delta_1$. In the self-consistency equations, we have the Cooper logarithm $l_\alpha$ which takes the form $l_a=\ln(2\Lambda/\Delta_0)\equiv l_0$ and  $l_b=l_c=\ln(2\Lambda/\Delta_1)\equiv l_1$. Observe that $l_0=l_1-\ln r$. Using these in Eq. (\ref{eq:self_con1}) we get
\bea\label{eq:self_con}
r&=&-2V_0\nu_F^1l_1\cos\phi,\nn\\
(1+V_1\nu_F^1l_1)\cos\phi&=&-rV_0\nu_F^0(l_1-\ln r),\nn\\
\Delta_1\sin\phi&=&\Delta_1V_1\nu_F^1l_1\sin\phi,
\eea
where $\nu_F^0$, $\nu_F^1$ are the densities of states in band $a$ and bands $b,c$, respectively. These three equations determine the parameters $r\in[0,\infty)$, $l_1>0$, and $\phi\in[0,\pi]$ subject to the condition  that $|\cos\phi|<1$. The parameter $l_1$ is only sensitive to the overall scale of the gap and is not so interesting to the present discussion.

\paragraph{The TRS phase:}
In this phase $\phi=0$ or $\pi$ and hence the last line in Eq. (\ref{eq:self_con}) is not enforced. The parameter $r$ is obtained by solving the transcendental equation \beq\label{eq:trans}(\nu_F^0/\nu_F^1)r^2+(2\nu_F^0V_0\ln r+V_1/V_0)rc-2=0,\eeq   
where $c=\cos\phi=\pm1$. Then, $\nu_F^1l_1=-rc/2V_0$. Since $l_1>0$ we have $c=-{\rm sgn}(V_0)$. That is, the relative sign between $\Delta_0$ and $\Delta_1$ ($s^{++}$ vs $s^\pm$) can be switched with the attractive or repulsive nature of the interband interaction $V_0$. 

To explore the phase diagram, consider the limit $V_0\rightarrow 0$ where we get $r=0$ implying $\Delta_0=0$, $\nu_F^1l_1=-1/V_1$, and $\phi=0$ or $\pi$. This requires $V_1<0$ and is the result of an effective 1 band model where $b,c$ are seen as one with intra-band interaction $V_1$. On the other hand, in the limit $V_1\rightarrow 0$ we have $\phi=0$ or $\pi$. We still have to solve Eq. (\ref{eq:trans}) to find $r$. In these two limits, we never get a solution where $\phi\neq\{0,\pi\}$, suggesting that to break TRS we need both $V_0$ and $V_1$ components, thereby enforcing necessity of 3 bands for having TRS broken state. As we change the parameters $V_0$ and $V_1$, this TRS phase continues all the way until $\nu_F^1l_1=1/V_1$. This is when the last line in Eq. (\ref{eq:self_con}) begins to be valid even for $\phi\neq 0$. This marks the onset of the TRS broken state.

\paragraph{The TRS broken phase:} In this phase, $\nu_F^1l_1=1/V_1>0$ is fixed and then we have
\bea\label{eq:sol1}
\cos\phi&=&-\frac{V_1}{2V_0}r,\nn\\
r&=&{\rm exp}\left(\frac{1}{\nu_F^1V_1}-\frac{V_1}{\nu_F^0V_0^2}\right),
\eea
subject to the condition that $|\cos\phi|<1$. A cross-section of the phase diagram across the TRS breaking transition is shown in Fig. \ref{phase-diag} which is obtained by keeping $V_1$ fixed and varying $V_0$. The distinction between $\nu_F^1$ and $\nu_F^0$ only introduces quantitative differences and thus for simplicity we set $\nu_F^1=\nu_F^0=\nu_F$ for all computations in this work. We will be interested in learning what the Raman response looks like across this phase transition.

\section{The results for Raman response} 
As outlined earlier, with $n=3$ we expect a total of $2n-1=5$ collective modes in our system. In the TRS side of the phase diagram, $n-1=2$ of them would be in the phase sector and hence visible in Raman spectroscopy. To figure out the spectral weights and characters of these modes we can deploy the general result from Sec. \ref{Sec:low-energy-model}. In Fig. \ref{fig:Contour} we plot the $\rm A_{1g}$ and $\rm B_{1g}$ responses across the TRS breaking phase transition. The $\rm A_{1g}$ spectrum of collective modes is fixed as soon the ground state is picked because it involves the same interactions and cannot change unless the ground state is changed. The spectral weights however, will depend on the Raman vertices. Figure \ref{fig:Contour} is shown for the choice $\gamma^b_{r}=-\gamma^c_{r}=0.5\gamma^a_{r}$ for $r\in\{\rm A_{1g},B_{1g}\}$. Although this is the first time such a spectrum has been explicitly calculated, the result is not surprising: a collective mode (labelled 1) softens at the TRS breaking transition. This mode bounces back and crosses another mode (labelled 2) that is induced in the TRS broken phase from the $2\Delta$ threshold. We plot the character of these modes in Fig. \ref{character} which shows that mode 1 is a pure phase one on the TRS side. It acquires an amplitude character as soon as it enters the TRS broken side and this character dominates deep inside the TRS broken state. Meanwhile, mode 2 emerges with largely an amplitude character and becomes phase-dominant deep inside the TRS broken phase. This mode will go on to soften with increasing $V_0$ leading to another transition boundary between the $s+is$ state and a different $s^{+-}$ state which has opposite phase of gaps between bands $b,c$. We don't show that transition here. There is more depth to these results and will be the subject of the discussion in following subsections.

The $\rm B_{1g}$ collective modes, on the other hand, depend on the choice of interactions in the $\rm B_{1g}$ channel, which are independent of the ground state. Figure \ref{fig:Contour} also shows the $\rm B_{1g}$ response for an interband driven scenario (more details in Sec. \ref{Sec:B1g}). It shows a BaSh mode on the TRS side that also has a tendency to soften and bounce back in the TRS broken side. Just like the $\rm A_{1g}$ response, there is another mode that is induced in the TRS broken phase near the $2\Delta$ threshold. What is striking here is that the BaSh mode displays a tendency to soften even though the $\rm B_{1g}$ interactions are held constant. This suggests that the approach towards TRS breaking state boosts both the competing sub-orders in the ${\rm A_{1g}}$ and ${\rm B_{1g}}$ channels. In fact, if the ${\rm B_{1g}}$ interaction in the TRS phase is strong enough, the approach towards the TRS breaking state can cause the BaSh mode to soften before the Leggett mode leading to an $s+id$ state. The type of $s+id$ ground state that would develop (i.e. the relative phase between various bands) depends on the phase character of the mode that softens. As we will discuss shortly, this phase character could be deduced from the spectral weights of the modes. In Fig. \ref{fig:Contour} we restricted ourselves to those ${\rm B_{1g}}$ interactions that are weak enough such that a transition to $s+id$ state is preceded by that to $s+is$ one. 

In the remainder of the text we will analyse the above spectrum and its variations to potentially deduce the sign of gaps in the bands, their interplay with the Raman vertices, and the sign of interband interaction, wherever possible. We will also illustrate the use of the recently discovered interaction induced selectivity of collective modes\cite{BenekLins2023} to make sense of the calculated spectral weights in the various scenarios. This rule, as a reminder, broadly associates the character of the collective modes to the Raman vertices and infers the relative spectral weights of the modes in the gap and near the $2\Delta$-threshold. For a pair of bands $a,b$, the rule suggests that a fluctuation in irrep $r$ with the form factor $\delta\phi^a_r\pm s\delta\phi^b_r$ couples to the Raman vertex with the form $\gamma^a_{r}\pm s\gamma^b_{r}$, respectively, where $s\equiv-{\rm sgn}[\Delta_a\Delta_bV_r^{ab}]$. Further, the mode $\delta\phi^a_r+s\delta\phi^b_r$ will be the lower energy mode and hence in the gap and $\delta\phi^a_r-s\delta\phi^b_r$ will be the higher energy mode which could be over the $2\Delta$ threshold and hence damped. 

\begin{figure*}[htb!]
\centering\captionsetup{justification=RaggedRight}
\includegraphics[width=1.0\linewidth]{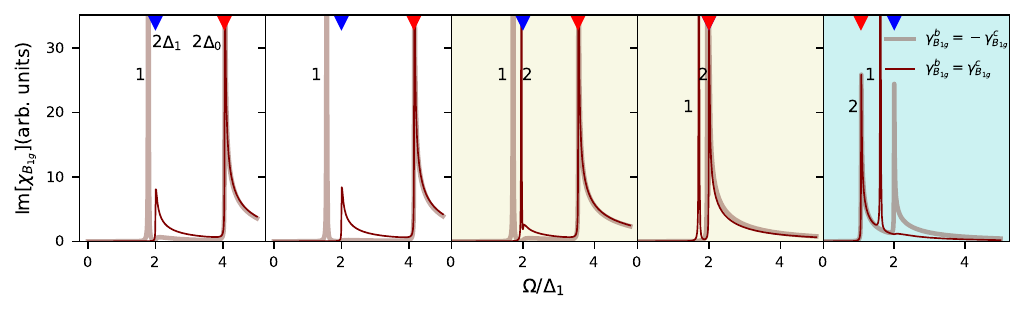}
\includegraphics[width=1.0\linewidth]{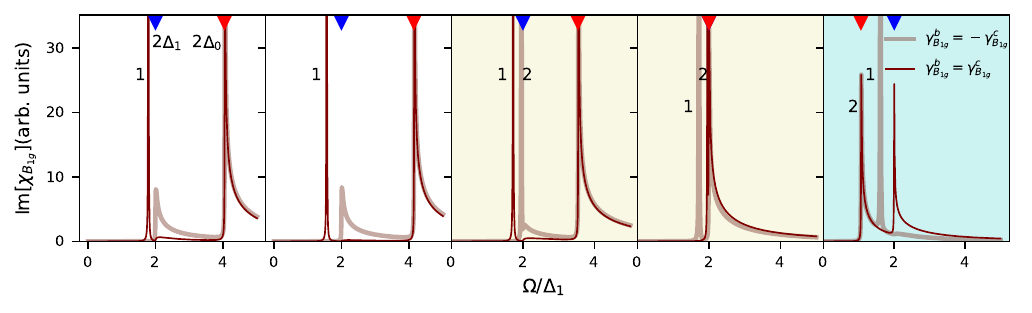}
\caption{The ${\rm B_{1g}}$ response, for $U_0=0$ and $\nu_FU_1=0.09$(top panel), $\nu_FU_1=-0.09$(bottom panel), across the $s^{++}\rightarrow s+is$ transition for different choices of Raman vertices. All panels follow the same convention as in Fig. \ref{fig:A1g} and are plotted for the same parameters. 
Observe that, in all cases, the spectral weights switch from the collective mode to the $2\Delta_1$ region or vice-versa after the TRS breaking transition. Since fluctuations from band $a$ are decoupled in this choice of interactions, the $2\Delta_0$ threshold peak always remains intact for different choice of Raman vertices and also across the TRS breaking transition.}
\label{fig:B1g1}
\end{figure*}
\begin{figure*}[htb!]
\centering\captionsetup{justification=RaggedRight}
\includegraphics[width=1.0\linewidth]{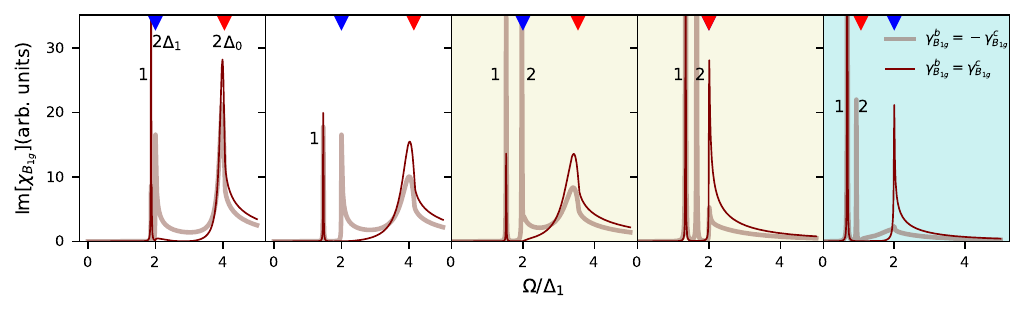}
\includegraphics[width=1.0\linewidth]{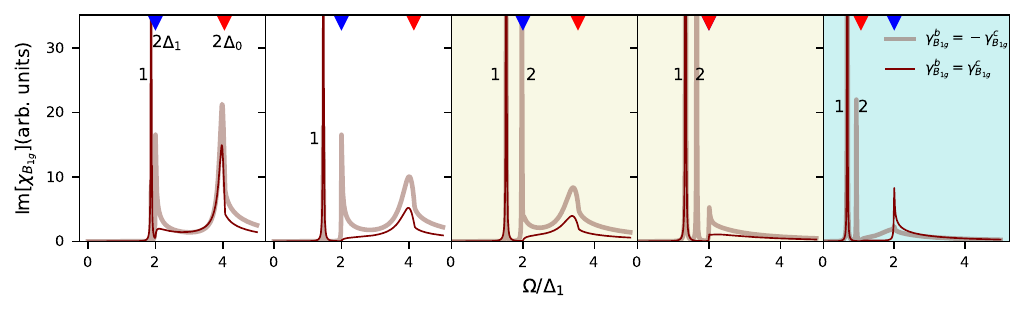}
\caption{The ${\rm B_{1g}}$ response, for $U_1=0$ and $\nu_FU_0=0.07$(top panel), $\nu_FU_0=-0.07$(bottom panel), across the $s^{++}\rightarrow s+is$ transition for different choices of Raman vertices. All other parameters are the same as in Fig. \ref{fig:B1g1}. The mode 1 includes phase character from band $a$ and thus couples to the Raman vertex which contains $\gamma^a_{\rm B_{1g}}\neq\pm\gamma^{b/c}_{\rm B_{1g}}$. That is why it has finite spectral weight for both the choices of the Raman vertices. The mode 2 inside the TRS broken phase involves phase fluctuations only from bands $b$ and $c$ and hence the selectivity is exact and this part of the spectrum is visible in only one choice of Raman vertex. The change in sign of $U_0$ does not affect the relative physics between band $b,c$ and hence the selectivity remains.}
\label{fig:B1g2}
\end{figure*}
\begin{figure*}[htb!]
\centering\captionsetup{justification=RaggedRight}
\includegraphics[width=1.0\linewidth]{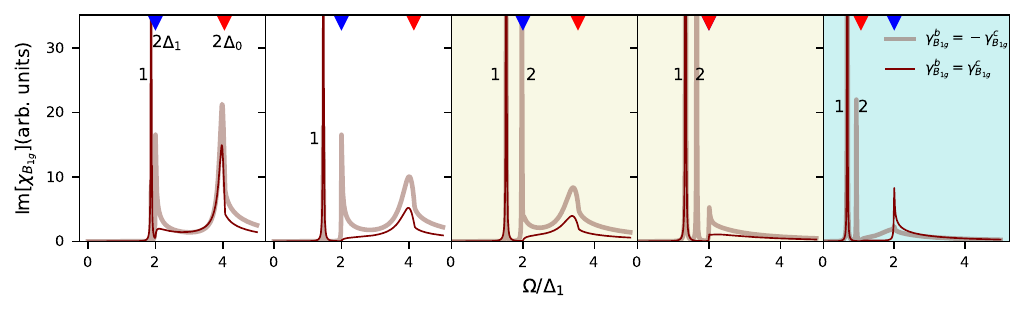}
\includegraphics[width=1.0\linewidth]{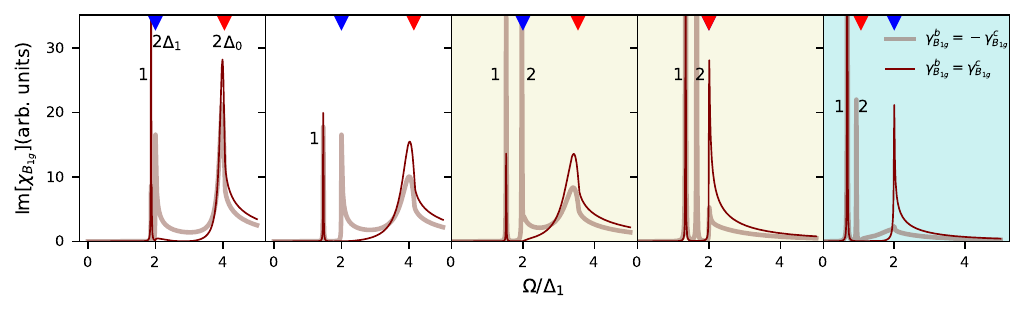}
\caption{The ${\rm B_{1g}}$ response, for $U_1=0$ and $\nu_FU_0=0.07$(top panel), $\nu_FU_0=-0.07$(bottom panel), across the $s^{+-}\rightarrow s+is$ transition for different choices of Raman vertices. All other parameters are the same as in Fig. \ref{fig:B1g2}. The only change from Fig. \ref{fig:B1g2} is that the $U_0>0$ and $U_0<0$ panels are switched, which is to be expected as the selection rule is sensitive to ${\rm sgn}[\Delta_a\Delta_{b/c}U_0]$.}
\label{fig:B1g2b}
\end{figure*}

\subsection{The eRS in the ${\rm A_{1g}}$ channel}\label{Sec:A1g}
In the $\rm A_{1g}$ irrep, due to the self consistency condition, we get $s=-{\rm sgn}[\Delta_a\Delta_bV_{\rm A_{1g}}^{ab}]=1$. This means that the in-phase mode ($\delta\phi^a_r+\delta\phi^b_r$) should be the low energy mode and couple to eRS as $\gamma^a_{\rm A_{1g}}+\gamma^b_{\rm A_{1g}}$. This is indeed what is expected of the BAG mode. Perhaps because this mode is renormalized up to the plasmon and hence not relevant in the low energy sector, it was missed in the literature that this mode would couple to the eRS spectrum via the $\gamma^a_{\rm A_{1g}}+\gamma^b_{\rm A_{1g}}$ combination of Raman vertices. The rule also implies that the out-of-phase mode ($\delta\phi^a_{\rm A_{1g}}-\delta\phi^b_{\rm A_{1g}}$) should be the higher energy mode and couple to eRS as  $\gamma^a_{\rm A_{1g}}-\gamma^b_{\rm A_{1g}}$. This is indeed the well known result for the 2-band Leggett mode\cite{Cea_prb16}. In fact, in our model where bands $b,c$ are identical, we can expect the Leggett mode involving bands $b,c$ to not couple  (go ``dark") when $\gamma_{\rm A_{1g}}^b=\gamma_{\rm A_{1g}}^c$. This is exactly what we see in Fig. \ref{fig:A1g} on the TRS side, where we plot the eRS spectrum for select values of $V_0$. This serves as an example for a case where there could be a phase transition to the $s+is$ without any visible softening of a Leggett mode.

To better understand why the rules work the way they do, let us evaluate these general expectations explicitly on the TRS side by substituting for the $\Delta_\alpha$'s and the $\Pi^\alpha_{ij}$'s in Eq. (\ref{eq:response_general}). Using the expressions for $l_\alpha$ from the self-consistency relations, we get the response to be
\bea\label{eq:TRS3bandResp}
\chi_{\rm A_{1g}}(\Omega)&=&[(\gamma^a_{\rm A_{1g}}-\gamma^b_{\rm A_{1g}})^2+(\gamma^a_{\rm A_{1g}}-\gamma^c_{\rm A_{1g}})^2]\frac{4\Delta_0\Delta_1/cV_0}{\Omega_1^2-\Omega^2}\nn\\
&&+(\gamma^b_{\rm A_{1g}}-\gamma^c_{\rm A_{1g}})^2\frac{(4\Delta_1^2/V_1)(\Omega_3^2-\Omega^2)}{(\Omega_1^2-\Omega^2)(\Omega_2^2-\Omega^2)},\nn\\
\text{where},\nn\\
\Omega_1^2&=&\left(\frac{1}{\nu_F^1I_1}+\frac2{\nu_F^0I_0}\right)\frac{2\Delta_0\Delta_1}{cV_0}\nn\\
\Omega_2^2&=&\frac{2(V_1\Delta_0\Delta_1+2cV_0\Delta^2_1)}{cV_0V_1\nu_F^1I_1}\nn\\
\Omega_3^2&=&\frac{2(V_1\Delta^2_0+2V_0c\Delta_1\Delta_0)}{V_0^2\nu_F^0I_0}.
\eea
Since the bands $b,c$ are identical, we will have $|\gamma^b_{\rm A_{1g}}|=|\gamma^c_{\rm A_{1g}}|$. If all three effective mass vertices were identical we would get a null response, which is expected as we would recover the total charge vertex which does not fluctuate due to charge conservation. Here we can note all the following relevant points : (i) All modes couple with $(\gamma_\alpha-\gamma_\beta)^2$ form factors, which explains why some modes can go dark if appropriate conditions on the vertices are met. (ii) We observe that the expression potentially hosts $n-1=2$ poles (with $n=3$). These are indicated by the `frequencies' $\Omega_1^2$ and $\Omega_2^2$ \footnote{These are actually transcendental equations in general which may or may not have solutions that correspond to long-lived modes.}. With only interband interactions, on the TRS side, we get one long lived Leggett mode in the gap and the other mode beyond the $2\Delta$ threshold. (iii) The self consistency equations enforce ${\rm sgn}(cV_0)=-1$. This causes the result to not be sensitive to the phase of the ground state ($s^{++}$ vs $s^{\pm}$) as $c$ and $V_0$ always come as a product.

Upon studying the character of the long lived mode we find that the fluctuating terms are indeed phase-like and such that phase terms from bands $b,c$ are opposite in sign to each other without involving band $a$(see Appendix \ref{Sec:App2}).  According to the rule such a mode should couple to effective Raman vertex $\gamma^b_{\rm A_{1g}}-\gamma^c_{\rm A_{1g}}$. This is why in Fig. \ref{fig:A1g} the Leggett mode only shows up for $\gamma^b_{\rm A_{1g}}=-\gamma^c_{\rm A_{1g}}$ and is dark for $\gamma^b_{\rm A_{1g}}=\gamma^c_{\rm A_{1g}}$.

In the TRS broken phase the general expressions become too unwieldy to show, but the ${\rm A_{1g}}$ response across the TRS breaking phase transition is plotted in Fig. \ref{fig:A1g} for two allowed cases of $\gamma_{A_{1g}}^b=\pm\gamma_{A_{1g}}^c$ for this model. The bounced-back mode (1) acquires an amplitude character with two characteristics (see  Appendix \ref{Sec:App2}): it mixes amplitude and phase characters; and it mixes contributions from band $a$. What is interesting is that the contribution from band $a$ only comes via the amplitude sector. The phase sector of the bounced-back mode is still exactly the same as it was on the TRS side. This is why the selectivity of the mode remains exact with respect to the choices of $\gamma_{\rm A_{1g}}^b\pm\gamma_{\rm A_{1g}}^c$ as is evident from the TRS broken panels of Fig. \ref{fig:A1g}. While the fact that there is no phase component of band $a$ is certainly an artifact of the model, this result, nevertheless, shows that the coupling of the collective modes to the eRS is entirely from the phase sector even in the TRS broken side. This is because if the phase sector of band $a$ were to be involved, then the vertex combinations $\gamma_{\rm A_{1g}}^a\pm\gamma_{\rm A_{1g}}^{b/c}$ would contribute. Since band $a$ is different from band $b,c$, the selection would not be exact. But we see in Fig. \ref{fig:A1g} an exact selectivity for the mode 1.

The second mode (2) that emerges from $2\Delta$ region also has mixed character from all the bands, but importantly involves the phase fluctuations from band $a$ (see Appendix \ref{Sec:App2}). According to the rule, we will now have contributions from $\gamma_{\rm A_{1g}}^a\pm\gamma_{\rm A_{1g}}^{b/c}$ vertices, and thus the spectral weight from this mode should be finite, but different, in the two choices of Raman vertices in Fig. \ref{fig:A1g}. This is indeed what the plot shows. Note also that the phase character of mode 2 with respect to bands $b,c$ is the opposite of mode 1. We will see below that this feature that the phase part of the fluctuations of a mode is what couples to eRS, remains true in every case we analyse.

It is also evident from Fig. \ref{fig:A1g} that the entry into the TRS broken phase is not marked with any characteristic change at frequencies beyond the $2|\Delta|$ region, although deep inside the TRS broken state, the spectrum above $2|\Delta|$ acquires sharper features. We think that this is probably a feature of this model and nothing generic. Finally, even in the TRS broken phase, just like the TRS phase, we verified explicitly that changing the phase of the ground state on the TRS side from $s^{++}$ to $s^{+-}$ (by changing $V_0\rightarrow-V_0$) did not alter the response.

\subsection{Raman response in the ${\rm B_{1g}}$ channel}\label{Sec:B1g}
The advantage of electronic Raman scattering is that the same experiment provides access to not just excitations in the $r={\rm A_{1g}}$ irrep but also in other irreps. We can take a look at $r={\rm B_{1g}}$ irrep to analyze how the spectrum could evolve across a TRS breaking transition. Studying this response requires no new calculations. The ground state remains the same as that in all the cases considered for $r={\rm A_{1g}}$. What is different here is that $V^{\rm pp}_{{\rm B_{1g}},\alpha\beta}$ is now a parameter that is independent of the ground state and thus can be anything. This largely increases the phase space of possibilities. But to lay within the scope of this work we can focus on the interband driven ${\rm B_{1g}}$ interactions modeled as
\beq\label{eq:int_b1g}
V^{\rm pp}_{\rm B_{1g}}\rightarrow\begin{pmatrix}
    0&U_0&U_0\\
U_0&0&U_1\\
U_0&U_1&0
\end{pmatrix}.
\eeq
We can now explore the contribution of each of the interband interactions. Note that in the limiting cases, we will consider below by setting $U_0$ or $U_1$ to zero, the ground state always remains coupled with the phase diagram shown in Fig. \ref{phase-diag}.

The discussion of spectral weights of the resulting modes in the $\rm B_{1g}$ channel will involve many cases because $s=-{\rm sgn}[\Delta_a\Delta_bV_{\rm B_{1g}}^{ab}]$ can switch sign based on the phase of the ground state or the attractive or repulsive nature of the $\rm B_{1g}$ interaction (compared to the $\rm A_{1g}$ case where $s=1$). We shall discuss them on a case-by-case basis. For definiteness, our ground state will be of the $s^{++}$ nature such that bands $a$ and $b,c$ are in phase on the TRS side. We will discuss the results for an $s^{+-}$ state (opposite phase between bands $a$ and $b,c$) whenever appropriate.

\subsubsection{No ${\rm B_{1g}}$ interactions}
In the limit $U_{0},U_1\rightarrow 0$ the response in the TRS and TRS broken phase is simply that of three independent bands which is the addition of $1/\sqrt{\Omega-2\Delta_0}$ and $1/\sqrt{\Omega-2\Delta_1}$ features. We don't discuss this further. If there are purely repulsive interactions such that there are no long-lived collective modes, we simply get some broad features near $2\Delta_1$ and $2\Delta_0$.

\subsubsection{One decoupled band}
Consider the limit $U_0\rightarrow 0$ but with finite $U_1$. The system's fluctuations are such that those from band $a$ (order parameter $\Delta_0$) are decoupled from the rest. In this case, we expect to get a non-interacting single band response from band $a$ leading to a $1/\sqrt{\Omega-2\Delta_0}$ feature. This feature is evident all across the transition in both panels (for repulsive and attractive $U_1$) in Fig. \ref{fig:B1g1}. Since this feature belongs to band $a$, this part of the response is not sensitive to $\gamma^b_{\rm B_{1g}}=\pm\gamma^c_{\rm B_{1g}}$. 

Added to this response will be that of a two (identical) band scenario where the fluctuations are coupled via interband interactions ($U_1$). For the bands $b,c$, which are always in phase in the TRS side, we get $s=-{\rm sgn}[\Delta_b\Delta_c U_1]=-{\rm sgn}[U_1]$. This means that the low energy mode (within the gap) should have the form factor $\delta\phi^b_{\rm B_{1g}}+s\delta\phi^c_{\rm B_{1g}}=\delta\phi^b_{\rm B_{1g}}-{\rm sgn}[U_1]\delta\phi^c_{\rm B_{1g}}$. This should couple to eRS as $\gamma_{\rm B_{1g}}^b-{\rm sgn}[U_1]\gamma_{\rm B_{1g}}^c$. So, for a repulsive interband interaction ($U_1>0$) we expect the lower energy BaSh mode to have an out-of-phase character suggesting that if it led to an $s+id$ ground state it would be one with opposite phases in bands $b$ and $c$. For an attractive $U_1$ interaction the in-phase mode becomes the lower energy mode. Further the Raman vertex that picks up the spectral weight would also different in each of these cases in accordance with the selection rule. All these points are exactly what we see in the two panels of Fig. \ref{fig:B1g1}. The character of the modes in each case are also presented in Appendix \ref{Sec:App2} for verification.

Upon entering the TRS broken phase, many interesting observations can be made. Staying with mode 1 for $\gamma_{\rm B_{1g}}^b=-\gamma_{\rm B_{1g}}^c$, we see that the selectivity is still maintained despite the mixing from the amplitude sector (see Appendix \ref{Sec:App2}). This further supports that even in the TRS broken phase only the phase fluctuations couple to the eRS spectrum. Mode 1 also loses spectral weight to the $2\Delta_1$ peak deep inside the TRS broken state. This is understood also from the selection rule. For identical bands with pure interband interactions, there are two modes, one attractive and one repulsive, with the in-phase mode coupling to $\gamma_{\rm B_{1g}}^b+\gamma_{\rm B_{1g}}^c$ and the out-of-phase to $\gamma_{\rm B_{1g}}^b-\gamma_{\rm B_{1g}}^c$\cite{BenekLins2023}. Since deep inside the TRS broken phase where $\Delta_b\approx-\Delta_c$, there is a switch in $s=-{\rm sgn}[\Delta_b\Delta_c U_1]$. This would switch the association of the spectral weights of the modes from one vertex to the other manifesting in the transfer of spectral weight from the collective mode to the $2\Delta_1$ peak for the vertex $\gamma_{\rm B_{1g}}^b=-\gamma_{\rm B_{1g}}^c$. This is exactly what is seen in Fig. \ref{fig:B1g1} in both panels. For the vertex $\gamma_{\rm B_{1g}}^b=\gamma_{\rm B_{1g}}^c$, the spectral weight actually moves to the collective mode.

Moving to mode 2, we observe that it is induced with a phase character opposite to that of mode 1(see Appendix \ref{Sec:App2}). This aspect is true for both panels and means that mode 2 should be visible in the Raman vertex combination opposite to that of mode 1. This complimentary nature of modes 1 and 2 is clearly evident from both panels in Fig. \ref{fig:B1g1}. 

It pointed out in Ref. \cite{BenekLins2023} that the response in $\rm B_{1g}$ irrep is sensitive to the sign of the gaps in the ground state. While this is true in general, in this case, the $s^{+-}$ phase in the TRS side would involve a sign change between bands $a$ and $b,c$ and not between $b$ and $c$. Thus, with $U_0=0$ and the band $a$ being decoupled from the rest, there would be no sensitivity to $s^{++}$/$s^{+-}$ states. We explicitly verified that this is the case. Nevertheless, note that deep in the TRS broken state where $\Delta_b\approx-\Delta_c$, we did find the response to change in a characteristic manner, in support of the above sensitivity claim.

\subsubsection{All coupled bands}
Consider the other limit of interband interactions were $U_1\rightarrow 0$ but with finite $U_0$, where the fluctuations of all bands are coupled. In this case, since $U_1=0$, the two bands $b,c$ effectively act as a unit. However, they do so in the following manner. Because they are simultaneously driven by band $a$, the effective interaction that would emerge between bands $b$ and $c$ is such that the intra- and interband interaction between these bands is the same (and repulsive). In such a system, the collective modes could be thought of as being derived from the interaction of band $a$ with the subsystem $b,c$. In the subsystem $b,c$ one can form and in-phase and an out-of-phase mode, with coupling tendency to $\gamma_{\rm B_{1g}}^b+\gamma_{\rm B_{1g}}^c$ and $\gamma_{\rm B_{1g}}^b-\gamma_{\rm B_{1g}}^c$, respectively. The out-of-phase mode remains decoupled while the in-phase mode couples to the fluctuations of band $a$ to give modes of the form $\delta\phi^a\pm\delta\phi^{b/c}$ which couple to eRS as $\gamma_{\rm B_{1g}}^a\pm\gamma_{\rm B_{1g}}^{b/c}$, respectively, where $\delta\phi^{b/c}\propto\delta\phi^b+\delta\phi^c$ and $\gamma_{\rm B_{1g}}^{b/c}\propto\gamma_{\rm B_{1g}}^b+\gamma_{\rm B_{1g}}^c$. As per the selection rule, $s=-{\rm sgn}[\Delta_a\Delta_{b}U_0]=-{\rm sgn}[U_0]$. Then, for the repulsive interaction we can claim that the $\delta\phi_a-\delta\phi_{b/c}$ mode would be the lower energy mode and the $\delta\phi_a+\delta\phi_{b/c}$ mode would be the higher energy. These two modes would be visible in both choices of the Raman vertices due to incomplete cancellations due to the presence of $\gamma_{\rm B_{1g}}^a$. The relative phase mode from $b,c$, however, would couple as $\gamma_{\rm B_{1g}}^b-\gamma_{\rm B_{1g}}^c$ and hence be selective. This is indeed the observation in Fig. \ref{fig:B1g2}. It so happens that in the chosen model the $\delta\phi_b-\delta\phi_c$ mode is $\approx 2\Delta_1$. The characters of the modes that are presented in Appendix \ref{Sec:App2} confirm the above expectations.

Upon entering the TRS broken side, the BaSh mode 1 in the gap turns back, retaining its character. And just like in the previous subsection, there is also an additional collective mode 2 that emerges from $2\Delta_1$ region. The character of this mode is such that it amplitude fluctuations from all bands contribute, but the phase ones do so only form bands $b,c$, that too in an out-of-phase manner (Appendix \ref{Sec:App2}). If only the phase sector of the mode is coupled to the eRS, then we should expect this part of the spectrum to only be visible for the vertex choice of $\gamma^b_{B_{1g}}-\gamma^{c}_{B_{1g}}$. This is indeed what is seen in Fig. \ref{fig:B1g2}.

Finally, we note that the above evolution was for the $s^{++}$ state. If we change the ground state on the TRS side to $s^{+-}$ then from the selectivity rule we expect that this should swap the evolution of $U_0>0$ with that of $U_0<0$. This swap is evident when Fig. \ref{fig:B1g2b} is compared with Fig. \ref{fig:B1g2}. 

\section{Conclusion}\label{Sec:Conclusion}
We have extended the theory of electronic Raman scattering for a general multiband singlet superconductor to include the case of TRS breaking and traced the response in the different irreps. We also prescribed a way to characterize the fluctuation form factors showing the equivalence between the Raman vertex and linear response kernel. We show that even when the TRS is broken, where the collective modes are expected to have a mixed amplitude and phase character, the coupling to eRS is only through the phase sector of the mode.

We demonstrated the gauge invariance of the formalism in the TRS phase, reproduced the usual softening of the Leggett mode in $\rm A_{1g}$, and showed that the TRS broken state is characterized by the introduction of another mode into the gap. While this has admixture from the amplitude sector, its phase character is opposite to the already existing Leggett mode. We also showed that there is simultaneous tendency for a BaSh mode to soften in the $\rm B_{1g}$ response, driven by $\rm A_{1g}$ interactions and the proximity to the TRS broken state, a feature that has already been reported in the context of Ba$_{1-x}$K$_x$Fe$_2$As$_2$\cite{Bohm2014,Bohm2018}.

While our formalism naturally calculates the spectral weights of all the modes, we also demonstrated the consistency of these weights with the interaction induced selection rule\cite{BenekLins2023}, pointing out certain cases where there could be dark Leggett and BaSh modes thereby triggering a TRS breaking phase transition without a visible softening of the collective mode. We even identified features in the spectrum that arise due to the sign change of the order parameter in the TRS broken phase.

These results provide a comprehensive view of the low energy collective modes in multiple symmetry channels within the linear response. Although the specific results discussed are for a 3-band model, the formalism is applicable to any number of bands. Knowing this formalism and the response, one is now in a better position to model systems with the chiral $d+id$, $p+ip$ states, or even the non-linear responses such as in the THz third harmonic generation experiments. These would be some future goals starting from this formalism.

\paragraph*{Acknowledgements.} We would like to acknowledge useful conversations with I. Benek-Lins. S.S. and S.M. were funded by the Natural Sciences and Engineering Research Council of Canada (NSERC) Grant No. RGPIN-2019-05486. S.S. was supported by the Horizon Postdoctoral Fellowship from Concordia University.

\appendix
\bwt
\section{Structure of $[\Pi^\alpha_{ij}]$}\label{Sec:App1}
Consider explicitly ${\rm Det}[\Pi^{\alpha}_{ij}]$, with $\Pi^\alpha_{11}\rightarrow\chi^\alpha_{11}\equiv\Pi^\alpha_{11}+2\nu_Fl_\alpha$, and $\Pi^\alpha_{22}\rightarrow\chi^\alpha_{22}\equiv\Pi^\alpha_{22}+2\nu_Fl_\alpha$ where $i,j\in\{1,2,3\}$. This is given by
\bea
{\rm Det}[\Pi^{\alpha}_{ij}]&=&\chi^\alpha_{11}(\chi^\alpha_{22}\Pi^\alpha_{33}-\Pi^\alpha_{32}\Pi^\alpha_{23})
-\Pi^\alpha_{12}(\Pi^\alpha_{21}\Pi^\alpha_{33}-\Pi^\alpha_{23}\Pi^\alpha_{31})
+\Pi^\alpha_{13}(\Pi^\alpha_{21}\Pi^\alpha_{32}-\chi^\alpha_{22}\Pi^\alpha_{31})\nn\\
&=&\chi^\alpha_{11}\chi^\alpha_{22}\Pi^\alpha_{33}+\chi^\alpha_{11}(\Pi^\alpha_{23})^2+(\Pi^\alpha_{13})^2\chi^\alpha_{22}-(\Pi^\alpha_{12})^2\Pi^\alpha_{33}-2\Pi^\alpha_{12}\Pi^\alpha_{23}\Pi^\alpha_{13}. 
\eea
Plugging the form of correlation functions $\Pi^\alpha_{ij}$ listed in Eq. (\ref{eq:pis}) we get
\bea
\frac{{\rm Det}[\Pi^{\alpha}_{ij}]}{\left(\mathscr{I}_\alpha(\Omega)/|\Delta_\alpha|^2\right)^3}&=&-\left[\frac{\Omega^2}{4}-\left(\Delta_\alpha^R\right)^2\right]\left[\frac{\Omega^2}{4}-\left(\Delta_\alpha^I\right)^2\right]\left[\left(\Delta_\alpha^R\right)^2+\left(\Delta_\alpha^I\right)^2\right]
-\left[\frac{\Omega^2}{4}-\left(\Delta_\alpha^R\right)^2\right]\frac{\Omega^2}{4}\left(\Delta_\alpha^R\right)^2\nn\\
&&+\frac{\Omega^2}{4}\left(\Delta_\alpha^I\right)^2\left[\frac{\Omega^2}{4}-\left(\Delta_\alpha^I\right)^2\right]-\left(\Delta_\alpha^R\right)^2\left(\Delta_\alpha^I\right)^2\left[\left(\Delta_\alpha^R\right)^2+\left(\Delta_\alpha^I\right)^2\right]+\frac{\Omega^2}{2}\left(\Delta_\alpha^R\right)^2\left(\Delta_\alpha^I\right)^2\nn\\
&=&0.
\eea
This zero is essential to ensure the presence of the BAG mode in the charge neutral theory, thereby ensuring gauge-invariance of the formalism.

\section{Band resolved character of the modes}\label{Sec:App2}
We solve the eigenvalue problem of Sec. \ref{Sec:Char} at the mode frequencies to find the eigenvector and from it deduce the weights of the amplitude and phase sectors. We show here the band resolved amplitude and phase breakdown of the modes in each irrep to the left and right of the TRS breaking transition. The results have the structure $\chi_m^r$ where, for mode $m$ in the $r^{\rm th}$ irrep, the first three entries correspond to the amplitude, phase and density components of band $a$, then the same for band $b$ and then the same for band $c$. For the $\rm A_{1g}$ case we have
\bea
[\chi_1^{\rm A_{1g}}]_{\nu_FV_0=-0.104}=
\begin{pmatrix}
0.00\\0.00\\0.00\\--\\0.00\\0.707\\0.00\\--\\0.00\\-0.707\\0.00
\end{pmatrix},~
[\chi_1^{\rm A_{1g}}]_{\nu_FV_0=-0.103}=
\begin{pmatrix}
-0.269\\0.00\\0.00\\--\\0.252\\0.632\\0.00\\--\\0.252\\-0.632\\0.00
\end{pmatrix},
[\chi_2^{\rm A_{1g}}]_{\nu_FV_0=-0.103}=
%
\begin{pmatrix}
0.00\\0.54\\0.00\\--\\-0.594\\0.019\\0.00\\--\\0.594\\0.019\\0.00
\end{pmatrix}.
\eea
We now show the band resolved character for various choices of the $B_{1g}$ interactions. For Fig.~\ref{fig:B1g1} and for $\nu_FU_1=0.09$ we get:
\bea
[\chi_1^{\rm B_{1g}}]_{\nu_FV_0=-0.104}=
\begin{pmatrix}
0.00\\0.00\\0.00\\--\\0.00\\0.707\\0.00\\--\\0.00\\-0.707\\0.00
\end{pmatrix},~
[\chi_1^{\rm B_{1g}}]_{\nu_FV_0=-0.103}=
\begin{pmatrix}
0.00\\0.00\\0.00\\--\\-0.387\\-0.591\\0.00\\--\\-0.387\\0.591\\0.00
\end{pmatrix},~
[\chi_2^{\rm B_{1g}}]_{\nu_FV_0=-0.103}=
\begin{pmatrix}
0.00\\0.00\\0.00\\--\\0.643\\-0.291\\0.00\\--\\-0.643\\-0.291\\0.00
\end{pmatrix},\nn\\
\eea

For Fig.~\ref{fig:B1g1} and for $\nu_FU_1=-0.09$ we get: 
\bea
[\chi_1^{\rm B_{1g}}]_{\nu_FV_0=-0.104}=
\begin{pmatrix}
0.00\\0.00\\0.00\\--\\0.00\\0.707\\0.00\\--\\0.00\\0.707\\0.00
\end{pmatrix},~
[\chi_1^{\rm B_{1g}}]_{\nu_FV_0=-0.103}=
\begin{pmatrix}
0.00\\0.00\\0.00\\--\\-0.387\\-0.591\\0.00\\--\\0.387\\-0.591\\0.00
\end{pmatrix},~
[\chi_2^{\rm B_{1g}}]_{\nu_FV_0=-0.103}=
\begin{pmatrix}
0.00\\0.00\\0.00\\--\\0.643\\-0.291\\0.00\\--\\0.643\\
0.291\\0.00
\end{pmatrix},\nn\\
\eea
For Fig.~\ref{fig:B1g2} and $\nu_FU_0=0.07$ we get:
\bea
[\chi_1^{\rm B_{1g}}]_{\nu_FV_0=-0.104}=
\begin{pmatrix}
0.00\\0.732\\0.00\\--\\0.00\\-0.481\\0.00\\--\\0.00\\-0.481\\0.00
\end{pmatrix},~
[\chi_1^{\rm B_{1g}}]_{\nu_FV_0=-0.103}=
\begin{pmatrix}
0.00\\0.723\\0.00\\--\\0.247\\0.420\\0.00\\--\\-0.247\\0.420\\0.00
\end{pmatrix},~
[\chi_2^{\rm B_{1g}}]_{\nu_FV_0=-0.103}=
\begin{pmatrix}
0.758\\0.00\\0.00\\--\\-0.397\\0.233\\0.00\\--\\-0.397\\-0.233\\0.00
\end{pmatrix},
\eea

For Fig.~\ref{fig:B1g2} and $\nu_FU_0=-0.07$ we get:
\bea
[\chi_1^{\rm B_{1g}}]_{\nu_FV_0=-0.104}=
\begin{pmatrix}
0.00\\0.732\\0.00\\--\\0.00\\0.481\\0.00\\--\\0.00\\0.481\\0.00
\end{pmatrix},~
[\chi_1^{\rm B_{1g}}]_{\nu_FV_0=-0.103}=
\begin{pmatrix}
0.00\\0.723\\0.00\\--\\-0.247\\-0.420\\0.00\\--\\0.247\\-0.420\\0.00
\end{pmatrix},~
[\chi_2^{\rm B_{1g}}]_{\nu_FV_0=-0.103}=
\begin{pmatrix}
0.758\\0.00\\0.00\\--\\0.397\\-0.233\\0.00\\--\\0.397\\0.233\\0.00
\end{pmatrix}.
\eea

\section{Other demonstrative examples of ${\rm B_{1g}}$ response}\label{App:Bash}
\begin{figure*}[htb!]
\centering\captionsetup{justification=RaggedRight}
\includegraphics[width=1.0\linewidth]{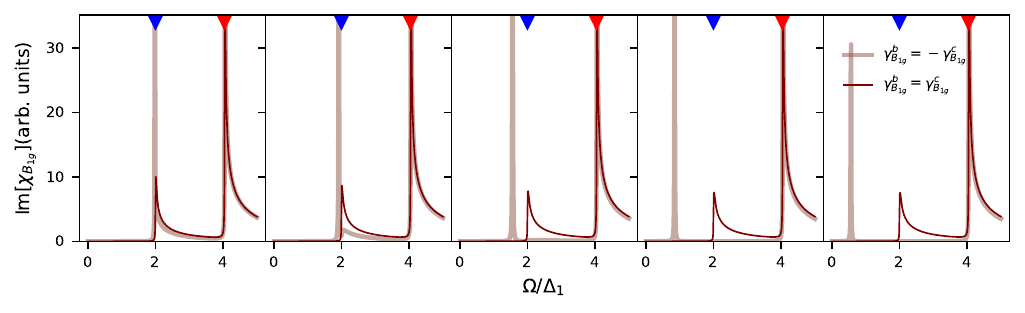}
\caption{${\rm B_{1g}}$ response: Bash mode soften as we increase the interaction $U_1$ in the $d-$wave channel. Here we choose  $U_0=0.00/\nu_F$, $V_0=-0.114/\nu_F$, $\nu_FU_1\in\{ 0.05,0.075, 0.1, 0.11, 0.112\}$, $\Delta_0/\Delta_1=2.02$.}
\label{AppFig}
\end{figure*}

Here we just demonstrate a usual BaSh mode behavior for a multiband system with increasing strength of interband  interaction. The BaSh mode softens as expected (see Fig. \ref{AppFig}).

\ewt
\bibliography{References.bib}
\end{document}